\documentclass[a4paper,twocolumn,11pt,unpublished]{quantumarticle}
\pdfoutput=1

\usepackage[utf8]{inputenc}
\usepackage[english]{babel}
\usepackage[T1]{fontenc}
\usepackage{graphicx}
\usepackage{dcolumn}
\usepackage{bm}
\usepackage{hyperref}

\usepackage{amssymb,amsmath}
\usepackage{amsthm}

\DeclareMathOperator{\csch}{csch}
\DeclareMathOperator{\sech}{sech}

\begin{document}

\title{Versatile Super-Sensitive Metrology Using \newline Induced Coherence}

\author{Nathaniel R. Miller}
\affiliation{University of Dayton, Department of Physics, Dayton, OH, 45469, United States}
\affiliation{Louisiana State University, Department of Physics and Astronomy, Baton Rouge, LA, 70803, United States}

\author{Sven Ramelow}
\affiliation{Faculty of Physics, Humboldt-University Berlin, Berlin 12489, Germany}
\author{William N. Plick}%
\affiliation{University of Dayton, Department of Physics, Dayton, OH, 45469, United States}

\begin{abstract}
\noindent We theoretically analyze the phase sensitivity of the Induced-Coherence (Mandel-Type) Interferometer, including the case where the sensitivity is ``boosted'' into the bright input regime with coherent-light seeding. We find scaling which reaches below the shot noise limit, even when seeding the spatial mode which does not interact with the sample \--- or when seeding the undetected mode. It is a hybrid of a linear and a non-linear (Yurke-Type) interferometer, and aside from the super-sensitivity, is distinguished from other systems by ``preferring'' an imbalance in the gains of the two non-linearities (with the second gain being optimal at \emph{low} values), and non-monotonic behavior of the sensitivity as a function of the gain of the second non-linearity. Furthermore, the setup allows use of subtracted intensity measurements, instead of direct (additive) or homodyne measurements \--- a significant practical advantage. Bright, super-sensitive phase estimation of an object with different light fields for interaction and detection is possible, with various potential applications, especially in cases where the sample may be sensitive to light, or is most interesting in frequency domains outside of what is easily detected, or when desiring bright-light phase estimation with sensitive/delicate detectors. We use an analysis in terms of general squeezing and discover that super-sensitivity occurs only in this case \--- that is, the effect is not present with the spontaneous-parametric-down-conversion approximation, which many previous analyses and experiments have focused on.          
\end{abstract} 

\maketitle


\section{\label{sec:level1}Introduction}

\noindent The ability to make precision measurements is paramount in science. It is also usually desirable to affect the system under study as little as possible. 

Classical interferometry, as typified by the Mach-Zehnder interferometer \cite{Zehnder1891, Mach1892}, splits light into two modes \--- one of which then interacts with the object to be measured \--- before they are recombined. The resulting interference pattern then provides phase information about the sample. The Mach-Zehnder interferometer (MZI) and its numerous variants typically rely on classical light to make measurements. However, this implies that the minimal detectable phase shift is never below the shot noise (or ``standard quantum'') limit of $\Delta \phi_{\mathrm{min}}^2=1/\langle N_{\mathrm{coh}}\rangle$, where $\Delta \phi_{\mathrm{min}}$ is the minimum detectable phase shift and $N_{\mathrm{coh}}$ is the average number of photons in the classical (coherent) field. Also, these systems most often use optical-frequency light as this regime is where the best detectors and optics are available. Detection in other domains is problematic \--- especially in the terahertz. 

\begin{figure}[h!]
\includegraphics[scale=.41]{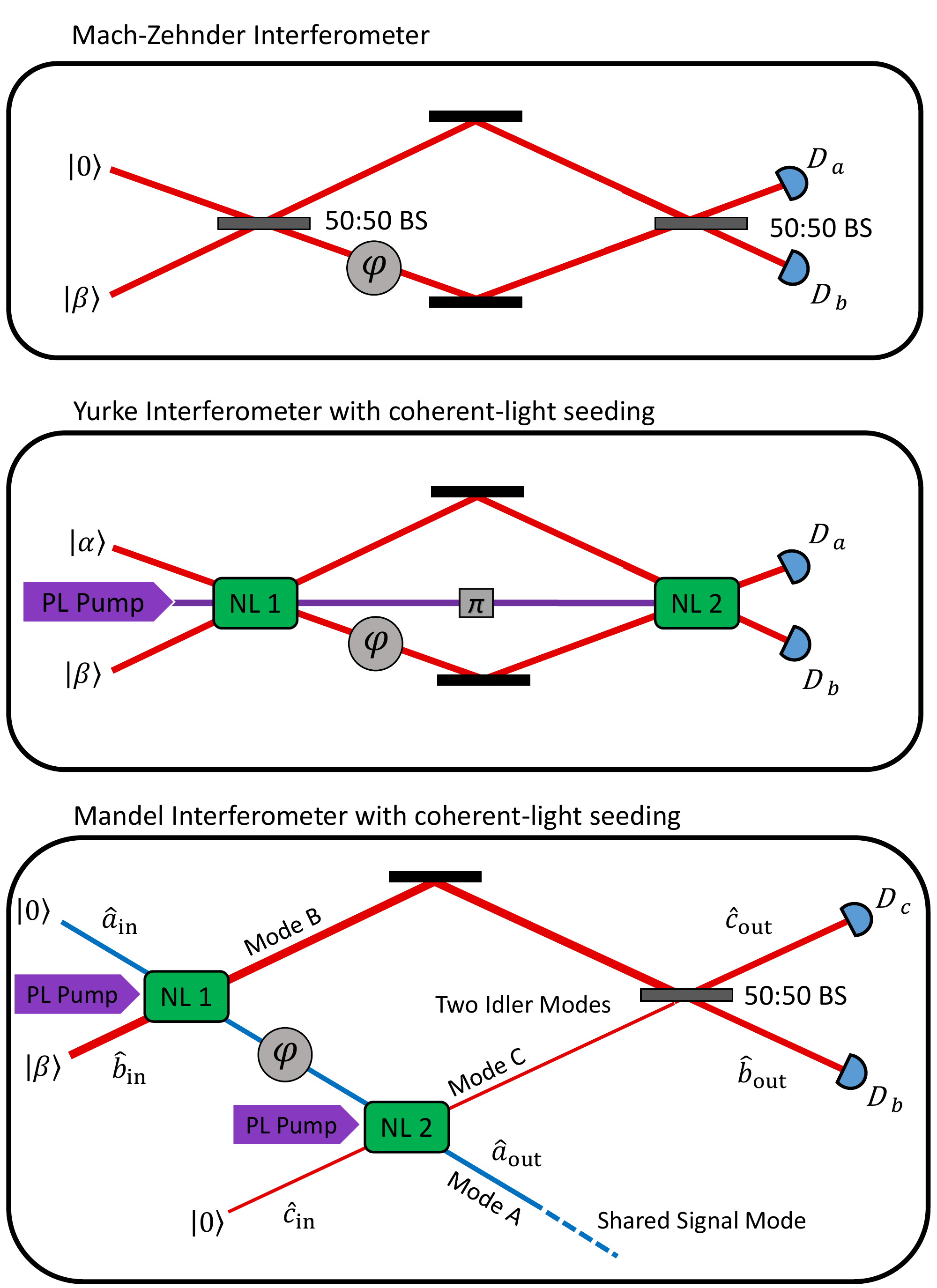}
\caption{Top: The Standard MZI; Middle: The Yurke or SU(1,1) Interferometer (seeded), Bottom: Our boosted Mandel-type setup, where two modes (operated on by $\hat{a}_{\mathrm{in}}$ and $\hat{b}_{\mathrm{in}}$, for the signal and first idler, respectively) are inputs into the first non-linearity (NL1 with squeezing parameter $r_{1}$) which is pumped with a laser which is also phase-locked to the pump of the second non-linearity (NL2 with squeezing parameter $r_{2}$). The inputs to the second non-linearity are the shared signal and second idler modes (operated on by $\hat{c}_{\mathrm{in}}$). Between NL1 and NL2 the signal mode interacts with an optical phase ($\varphi $), and after NL2 the mode is discarded. The idler modes are then mixed on a 50:50 beam-splitter, erasing the ``which-nonlinearity'' information before detection at $D_{b}$ and $D_{c}$. (Note that we use the opposite convention for signal/idler from Ref.\cite{Lemos2014}.) Though any of the three input modes could be boosted (seeded) with coherent light we examine, in the most detail, coherent-light-injection into the first idler mode ($|\beta\rangle $), as this is light which will not interact with the phase-imprinting sample, and which significantly improves sensitivity \--- as we will see.}
\label{fig:setup}
\end{figure}

The Mach-Zehnder Interferometer, along with two other types (including the one we study here), are presented in Figure \ref{fig:setup}.

Mitigating the limitations of shot noise is a vigorously-pursued avenue of research starting with the landmark paper of Carlton Caves \cite{Caves1981} where it was shown that so-called ``squeezed vacuum'' light could reduce the minimum detectable phase shift below the shot noise limit (SNL) in a MZI, when injected into the ``empty'' port of the device \--- the reason being that, unlike ``normal'' vacuum, squeezed vacuum carries phase information. This approach has even recently been applied to GEOS \--- a gravitational-wave detector \cite{LIGO2013}.  

Measuring a phase with higher precision than is possible with classical light has also been studied and experimentally explored in the frame-work of photon-number entangled states. The most prominent example of such states are the so-called ``N00N States'' \cite{N00N}. These can be written in the number basis as $|N\rangle_{a} |0\rangle_{b} +|0\rangle_{a} |N\rangle_{b}$ (hence the name), where the subscripts on the kets indicate interferometric path after the initial beam splitter in an MZI . In the ideal, but nonphysical, situation of zero loss these states can realize a Heisenberg scaling for the phase-sensitivity \--- that is a scaling that is proportional to $N$, where as classically only the $\sqrt{N}$ of the SQL is achievable. However, in the presence of loss, this scaling advantage quickly vanishes, as the absorption (detection) of a single photon gives complete path information and collapses the state. Another disadvantage of this scheme is that the efficient experimental generation of N00N states with more than 10 photons ($N=10$) is still an extraordinary challenge, far away from being relevant for practical phase-sensing tasks.

It is even possible to perform metrology with \emph{thermal} light in an MZI. However the scheme does indeed require a quantum resource \--- photon subtraction, which is a highly-non-classical operation. For further details see Ref.\cite{thermal}.  

Another avenue of research involves replacing the beam splitters of the MZI with nonlinear media ($\beta$-Barium Borate, etc.) These ``non-linear'' or ``entangled'' interferometers (see Refs.\cite{Chekhova2016}, and \cite{quantum entangled} and references therein) can push the sensitivity limit down to $\Delta \phi_{\mathrm{min}}^2=1/\langle N_{\mathrm{nl}}\rangle(\langle N_{\mathrm{nl}}\rangle+2)$, where $N_{\mathrm{nl}}$ is the average number of photons produced by the non-linearities. Such a setup is shown in the middle diagram in Figure \ref{fig:setup}. This shows a large improvement in scaling over the Michelson interferometer and was first discovered by Yurke, McCall, and Klauder. \cite{Yurke1985}. Hereafter we refer to this as a ``Yurke-Type'' interferometer. However, photon production by a two-mode squeezing non-linearity goes as $\langle\hat{N}_{\mathrm{nl}}\rangle=2\sinh^{2}(r)$, therefore since photon numbers are extremely low relative to other sources (such as a laser) the power of such setups (without further modification) is limited.  

\begin{figure}[h!]
\includegraphics[scale=.45]{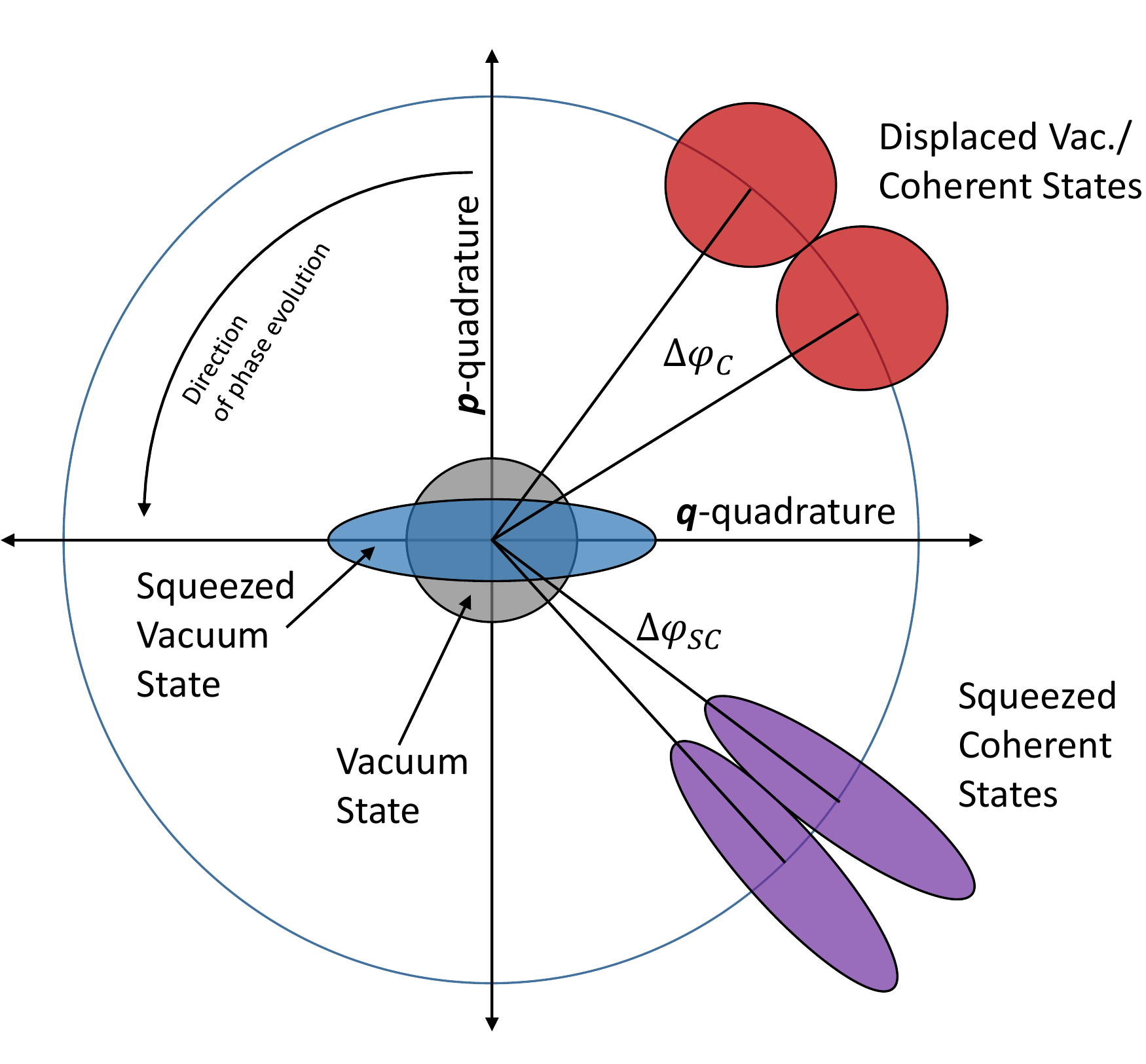}
\caption{Quadrature diagram in which several quantum-optical states are living. Vacuum states (grey) have no phase information (they are rotationally-symmetric and do not change as phase evolves). Coherent (or displaced vacuum \--- red) contain phase information proportionate to their displacement from the origin. That is, we can distinguish coherent states when the angle between them is $\Delta\phi_{C}$, which increases as intensity increases and states are displaced further. These are the states that traditional MZIs use. Squeezed states (blue) change their orientation as the phase evolves due to the asymmetry introduced by the squeezing. These are the states that are used in the Caves setup and in the original (unseeded) Yurke interferometer. Coherently-seeded non-linear interferometers provide the greatest potential advantage as they utilize squeezed coherent states (purple), which can be distinguished when the angle between them is $\Delta\phi_{SC}$. This angle is smaller since the states are displaced in the same way as the coherent states, but they are narrower in the phase direction.}
\label{quad}
\end{figure}

The physical mechanism for these phase-sensitivity improvements is illustrated in Figure \ref{quad}, where we show various states in quadrature space. Quadrature space is defined in terms of the $p$ and $q$ quadratures of the EM field as described in quantum optics. Light fields are defined both by the position in this space and by the variance. These diagrams can be thought of as a horizontal slice through the Wigner distribution.

The sensitivity can be pushed farther by seeding the modes of the original nonlinear media with coherent light \cite{Plick2010}, which has been further theoretically investigated \cite{Marino2012,Ou2012,Ma2018,Giese2017,Manceau2017,Florez2018,You2018} and implemented experimentally \cite{Jing2011,Hudelist2014,Lukens2016,Anderson2017,Manceau2017b,Anderson2017b,Chen2015} including several variations \--- accounting for loss and other experimental imperfections. Unlike the MZI the Yurke-type interferometers need not produce light modes of the same frequency. Indeed it can be stretched as far as to have one mode be optical light, and the other an atomic spin wave \cite{Chen2015}. The result is a new minimal detectable phase shift of $\Delta \phi_{\mathrm{min}}^2=1/\langle N_{\mathrm{nl}}\rangle\langle N_{\mathrm{coh}}\rangle(\langle N_{\mathrm{nl}}\rangle+2)$, where $N_{\mathrm{coh}}$ is now light from the seeded inputs. 

Several other interesting variants have recently been proposed, such as the so-called ``Pumped-up'' SU(1,1) interferometry where the pump beam is also utilized during measurement \cite{pumped}, and a scheme where an SU(2) interferometer is nested inside of an SU(1,1) \cite{su(2)}. Another recent paper has studied the use of optical parametric amplifiers placed into the individual arms of an MZI \cite{quantum interferometer}. All three of these schemes have yielded promising initial results.   

We propose a new variation on this successful setup based on the phenomenon of ``induced coherence'', an effect first discovered by Zou, Wang, and Mandel \cite{Zou1991} (hereafter a ``Mandel-Type'' interferometer) and brought to recent prominence with the imaging experiment of Lemos et al. which created phase images of an object in a light field which had never interacted with the object \cite{Lemos2014,Lahiri2015}. In these experiments, one mode (say, the signal) of the first non-linearity is seeded into the same mode of the second non-linearity, which \emph{induces coherence} between the idler modes. Phase information of an object in the shared signal mode is imprinted on the idler mode which is then mixed with the same mode from the first non-linearity on a standard beam splitter. This experiment can be thought of a type of quantum eraser (Ref.\cite{Ma2016} and references therein) where \emph{welcher-weg} (which-path, or perhaps more accurately ``which non-linearity'') information is erased by the alignment of the signal modes of the two non-linearities in conjunction with the beam splitter which mixes the idler modes. In other words, after the second non-linearity information could in-principle be obtained by seeing which idler mode contained light, but after the beam splitter this information is erased. Then the light field that passes through the sample is discarded. See the bottom diagram in Figure \ref{fig:setup}. This constitutes a type of hybrid interferometer described by neither the SU(2) formalism (standard MZI and variants), nor the SU(1,1) formalism (Yurke-Type interferometers). 

Such setups can even be used to perform tomographic bio-imaging \cite{Valles2018}. Another major potential application of this technique is spectroscopy, since samples may be probed in frequency domains which are different from where detection is performed \cite{Dinani2016,Paterova2018,Kalashnikov2016,Whittaker2017}. This has also been done with frequency combs and induced coherence \cite{Lee2018, Lee2019}. The Mandel-Type interferometer has also been used to perform optical coherence tomography \cite{frequency} and microscopy \cite{microscopy}. An equivalent device has also been implemented and studied in a microwave super-conducting cavity with application to continuous-variable quantum computing \cite{coherence, entanglement}. It is possible that these systems could also be modified in the way we describe below to realize the beneficial properties we present. In short, there is the potential for the results we describe to be applied to a wide variety of systems for use in several different applications. 

We study the injection of a coherent seed into the Mandel-Type device, allowing the sensitivity to be ``boosted'' into the bright super-sensitive regime. Also, the current work represents the first time such an interferometer has been studied specifically for phase sensitivity in metrology.

From the perspective of optics, this device would work as a Mandel-type interferometer with light fields displaced coherently in quadrature space, causing a more-noticeable interaction with the phase to be probed. One could think of how this device operates on the input optical fields in order to get a physical intuition. From this perspective, the first interaction (NL1 in Fig.\ref{fig:setup}) would classically be described as a relatively simple parametric amplifier \--- leading to an amplification of the seed mode (here either the vacuum or a coherent state such as from a phase-locked laser), and a conjugate copy of the seed in the second output. However, light is not a wave (or not \emph{entirely} one) so this approximation has to break down when the mean number of photons in the seed is sufficiently small and/or the gain is sufficiently large. Then the more-general description in terms of a Bogoliubov transformations leads to the by-now, well-known phenomenon of squeezed coherent output states with quadrature-entanglement between the two output modes. 

In our device one of those outputs is used to measure the unknown phase and then used as a seed in the second non-linear amplification step with a potentially different gain (but pumped coherently with respect to the first interaction). The output (idler) which is conjugate to the sensing field is afterwards interfered with the seeded idler from the first interaction. This is akin to a familiar homodyne measurement with a local oscillator (LO) \--- the crucial difference being that the LO here is not a coherent state, but still relatively strong (to boost the sensitivity), and phase locked with the idler output of the second interaction, allowing measurement of the unknown phase. 

What we find is that the device does indeed exhibit bright super-sensitivity. However, unlike the case of the standard Yurke setup, the second non-linearity does not need to be pumped as strongly as the first. In fact, after a certain pumping value, no further increase in sensitivity is obtained when considering direct intensity measurements. This will be of practical importance when designing experiments and real-world measuring devices. 

It is also the case that intensity-subtraction may be used instead of direct (total intensity) detection, or homodyne. Intensity subtraction is a much-more stable and straightforward procedure since noise which is shared between the modes (such as a vibration of the device in total) is cancelled out. 

Furthermore these effects are present even when the seeded mode is \emph{not} the mode that passes through the phase, meaning that bright light may be used to perform measurements on a sample (and also form images) without having that bright light shine through the sample. This has potential application to light-sensitive systems which are of strong interest in chemistry and biology. 

Additionally, the interacting mode need not be in the same frequency domain as the detecting mode, so the sample may be interrogated with one frequency and detection performed in another. 

Conversely the undetected mode may be seeded, yielding a similar increase in sensitivity without exposing the detectors to bright light.         

In Section II we theoretically describe the device in question, in Section III we analyze the phase sensitivity in a number of regimes, and in Section IV we compare this device to a several others and comment on why the first-quantized description is inadequate. In Section V we summarize and conclude.    

\section{\label{sec:level1}Theoretical Description}

\noindent In 2014 Lemos et al. \cite{Lemos2014} showed that the Induced Coherence experiment originally done by Zou et al. \cite{Zou1991} could be used to measure phase images in an undetected beam. 

Here, we extend the study of these systems by ``boosting'' \--- seeding the non-linearities with coherent light and calculating the minimum detectable phase shift as a function of the available parameters. Previous work has shown that boosting nonlinear medias with coherent light in interferometric setups can drastically increase the phase sensitivity \cite{Plick2010}. A diagram of the experimental setup is shown in the bottom diagram of Figure \ref{fig:setup}, which can be modeled by representing the output operators in terms of the input operators under the transformations 

\begin{align}
\mathbf{S_1 '}=
\begin{bmatrix}
	\mu_1 	& 	0 	& 	0 	& 	-\nu_1 	\\
    0 	& 	\mu_1 	& 	-\nu_1^*  	& 	0 	\\
    0 	& 	-\nu_1 	& 	\mu_1 &	0	\\
    -\nu_1^*	&	0	&	0	&	\mu_1	\\
\end{bmatrix} \,
\mathbf{\Phi '} =
\begin{bmatrix}
	e^{i\,\phi} 	& 	0 \\
0 	& 	e^{-i\,\phi} \\
\end{bmatrix},\hspace{3cm}\nonumber 
\\
\mathbf{S_2}=
\begin{bmatrix}
	\mu_2			 	& 	0 					& 	0 	& 	0 	& 	0 					& 	-\nu_2				\\
    0 					& 	\mu_2			 	& 	0  	& 	0 	& 	-\nu_2^*			& 	0					\\
    0 					& 	0 					& 	1	&	0	&	0					&	0					\\
    0					&	0					&	0	&	1	&	0					&	0					\\
    0					&	-\nu_2				&	0	&	0	&	\mu_2				&	0					\\
    -\nu_2^*			&	0					&	0	&	0	&	0					&	\mu_2				\\
\end{bmatrix},\nonumber \hspace{3cm}\quad
\\
\mathbf{A_{\mathrm{in}}}=
\begin{bmatrix}
	\hat{a}_{\mathrm{in}}	\\
    \hat{a}_{\mathrm{in}}^\dag\\
    \hat{b}_{\mathrm{in}}	\\
    \hat{b}_{\mathrm{in}}^\dag\\
    \hat{c}_{\mathrm{in}}	\\
    \hat{c}_{\mathrm{in}}^\dag\\
\end{bmatrix},\nonumber
\mathbf{BS '}=\frac{1}{\sqrt{2}}
\begin{bmatrix}
1	&	0					&	i	&	0					\\
0					&	1	&	0					&	-i	\\
i	&	0					&	1	&	0					\\
0					&	-i	&	0					&	1	\\
\end{bmatrix}.\hspace{4cm}\nonumber
\end{align}

\noindent And $\mathbf{\Phi}=\mathbf{\Phi '}\oplus \mathbf{I_{4}}$, $\mathbf{S_{1}}=\mathbf{S_{1} '}\oplus \mathbf{I_{2}}$, $\mathbf{BS}=\mathbf{I_{2}}\oplus \mathbf{BS '}$. The $\mathbf{I}$'s are the identity matrices of the dimension indicated by the subscript, and also $\mu_i=\cosh{r_i}$, $\nu_i=e^{i\psi_i}\sinh{r_i}$, with $r_i$ representing the squeezing parameter of the $i$-th non-linearity, and $\psi_{i}$ the phase. $\mathbf{BS}$ represents the beam splitter, $\mathbf{S_{1}}$ the first squeezer (non-linearity), $\mathbf{S_{2}}$ the second, and $\mathbf{\Phi}$ the phase to be probed. $\mathbf{A_{in}}$ is a vector composed of all the operators on which the transformation matrices act \--- the input operators. The final (output) operators can be found by applying the transformation 

\begin{align}
\mathbf{A_{out}}=\mathbf{BS\cdot S_2\cdot \Phi_1\cdot S_1\cdot A_{in}}.
\end{align} 

\noindent Then any detection operator at output can be written in terms of input operators working on the initial states, which we take to be coherent states (eigenstates of the annihilation operator, displaced vacuum which becomes vacuum in the limit of zero displacement). The reason to propagate the operators ``backwards'', rather than the state ``forwards'' through the device, is because in this case state propagation is much more cumbersome \--- even more so than what is employed here. Furthermore, the action of the output operators in terms of those output states would themselves not have a simple form. It is much more efficient to employ the analysis we use here.    

The output detector we study in the most detail is $D_{b}$. That is because the field going to this detector mixes with another mode before detection relative to the probe phase, and when it is boosted by the injection of coherent light optimal results are obtained in most cases. Due to the symmetric mixing action of the beam splitter, detection at detector $D_{c}$ will be similar, up to some relative phase. 

We note that injecting light into the other two ports is possible \--- and beneficial in some situations, as we will discuss.

\section{\label{sec:level1}Analysis of Phase Sensitivity}

We want to find the minimum detectable phase shift of the probe phase ($\phi $) which is given by 

\begin{align}
\Delta\phi_{\mathrm{min}}^2=\frac{\Delta\hat{O}^2}{|\partial_{\mathrm{\phi}}\langle\hat{O}\rangle |^2},
\end{align} 

\noindent as found in Ref.\cite{Braunstein1994}, for example. Where $\hat{O}$ is some general detection operator and the variance squared is given by 

\begin{align}
\Delta\hat{O}^2=\langle\hat{O}^{2}\rangle-\langle\hat{O}\rangle^{2}.
\end{align} 

In order to perform such calculations we need to find the first and second moments of the final detection operators. These calculations are straight-forward but extremely lengthy, therefore we created a program in the symbolic calculation software Mathematica$^\mathrm{TM}$ to perform them, using the NCAlgebra package \cite{NCA}.

The program (a simple, commented version is available in the supplementary materials or by request) proceeds by performing the relevant transformations. 

We will consider intensity detection at each output: $\hat{I}_A =  \hat{a}_{\mathrm{out}}^\dag \hat{a}_{\mathrm{out}}$, $\hat{I}_B =  \hat{b}_{\mathrm{out}}^\dag \hat{b}_{\mathrm{out}}$, and $\hat{I}_C =  \hat{c}_{\mathrm{out}}^\dag \hat{c}_{\mathrm{out}}$, as well as the difference (subtraction) intensity operators defined as $\hat{S}_{ij}=\hat{I}_{i}-\hat{I}_{j}$, where subscripts index the modes.

Qualitatively, the calculation can be described as follows: the first step is find the the output operators in terms of the input operators. That is, in the form $\hat{a}_{\mathrm{out}}=f_{\hat{a}}(\hat{a},\hat{a}^{\dagger},\hat{b},\hat{b}^{\dagger},\hat{c},\hat{c}^{\dagger})$, etc., where the operators in the parenthesis are the creation/annihilation operators of those modes at input. These are obtained by multiplication of all of the matrices representing the action of the device as in Eq.(1). 

The moments of these operators, needed for the calculation of the minimum phase shift, work on the state at input, which for the case of coherent input into mode $b$ can be written as $|0,\beta,0\rangle$. These are then eigenstates of the operator moments and the action of these operators is therefore straightforward, i.e. $\hat{a}|\alpha\rangle=\alpha|\alpha\rangle$, with $\hat{a}|0\rangle=0$ as a special case.

However what remains is to order the operators such that all daggered operators are on the left and all un-daggered operators are on the right. Since each operator is in-principle a function of six other operators, and since the second moments involve multiplying four such operators; this stage of the calculation is highly non-trivial. Thus the Mathematica program analytically performs all the commutations using $[\hat{a},\hat{a}^{\dagger}]=1$ etc., acts them on the states, and then simplifies the output equation as much as possible. We show an example of one of these calculations in a Mathematica notebook in a separate supplementary material (or by request). 

Strictly speaking, in order to achieve maximum sensitivity, there should be as much light seeding each of the detected modes as possible. However, it is more reasonable to consider a finite ``light budget''. In this case it's most advantageous to inject into mode $B$ since it does not pass though the sample and has as significant effect on sensitivity as mode $C$. Mode $A$ is less effective (for direct intensity measurement, for intensity subtraction they are equal \--- see later in this section) and must pass through the sample. Although if it is desirable to have light shine through the sample and \emph{not} reach the detectors (if the detectors are very fragile for example), then the sensitivity can be boosted in this way. A novel feature of the setup we propose here. 

The other key parameters are the squeezing values, and the sensitivity as a function of them can behave in several different ways depending on the values of the other parameters and which detector is monitored. We will focus on the case of greatest interest \--- when only mode $B$ is boosted with a coherent seed, phases are set to zero, and intensity measurements are also made in mode $B$. 

\begin{figure}[h!]
\includegraphics[scale=1.1]{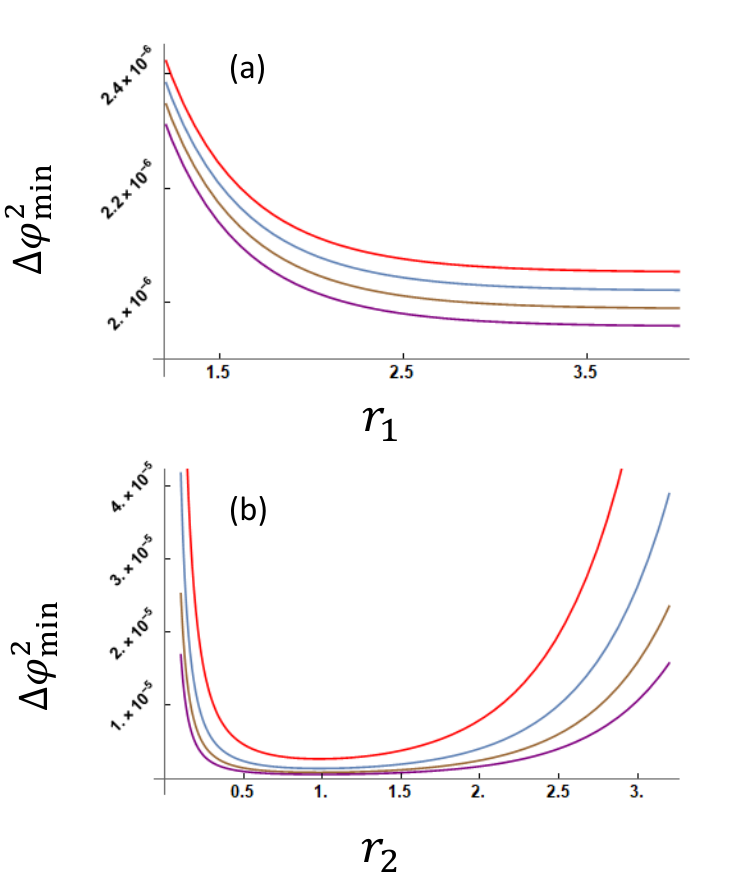}
\caption{The minimum detectable phase shift squared, as a function of each squeezing value for four values of the coherent input to mode B and the other squeezing set to one. In (a) $\beta_{\mathrm{red}}=1000$, $\beta_{\mathrm{blue}}=1008$, $\beta_{\mathrm{brown}}=1016$, and $\beta_{\mathrm{purple}}=1024$. In (b) $\beta_{\mathrm{red}}=1000$, $\beta_{\mathrm{blue}}=1400$, $\beta_{\mathrm{brown}}=1800$, and $\beta_{\mathrm{purple}}=2200$. All other parameters set to zero. The detection operator is the direct intensity at mode B.}
\label{squee2}
\end{figure}

Figure \ref{squee2} shows the minimum detectable phase shift as a function of the two squeezing parameters.  

There are a couple features worth noting. 

Firstly, the dependence of $\Delta\phi_{\mathrm{min}}^{2}$ on $r_{2}$ has a U-shaped valley, with some optimal value of the squeezing. This is a behavior totally unlike the SU(1,1) interferometer \--- where an unbalancing favoring \emph{higher} gain is found \cite{Giese2017,Manceau2017}. This is likely a significant practical advantage as higher gains are more difficult to achieve, especially when exacting alignment of different optical modes is required. However, it is fair to note that the scheme in Refs.\cite{Giese2017,Manceau2017} is very advantageous in the case of loss. We do not study loss here, though it is an avenue we are pursuing.

Secondly, the squeezing in the first crystal has an outsize influence on the minimum detectable phase shift compared to the squeezing in the second crystal. We conjecture that the second non-linearity is only needed to induce the coherence between the optical modes, further squeezing of the vacuum does not aid in sensitivity since that mode never again interacts (or becomes coherent with) the measured modes, and likely merely adds noise. This effect is a significant practical advantage for experimental implementation, as one only needs to try and ``push'' the squeezing in the first non-linearity.      

Related to this second feature, while this manuscript was under review we became aware of some work on a similar system by Kolobov, Giese, Lemieux, Fickler, and Boyd \cite{new}. In that paper the authors theoretically study a Mandel-type interferometer with a variable-transmittance beam splitter inserted between the two non-linearities (where we have our optical phase), with different pump intensities for them as well. They look at the signal-to-noise, and visibility, of the output as a function of the gain in the non-linearities and the transmittance of the aforementioned beam splitter, paying special attention to the qualitative and quantitative difference between various regimes of gain. They also find that the visibility of the system may be optimized by proper choice of the transmittance of the beam splitter. This is likely related or linked to the effect we observe with respect to the \emph{sensitivity} as a function of $r_{2}$ \--- seen from a different perspective. We quote their explanation of the effect here (from the last paragraph of section 3 of Ref.\cite{new}):

``...due to seeding crystal A, crystal B produces more photons then the first one. Hence, both arms become unevenly populated and the visibility drops for increasing gain. But if the transmittance is sufficiently small, the seeding effect is suppressed (since most photons from crystal A are not transmitted) and both arms have roughly the same intensity. This explains why the visibility follows the coherence for a small transmittance and diminishes for a large one.'' Where their crystal A is our NL1, and crystal B is NL2. We believe this presents a nice complementary explanation to ours. We again note, that at least for the sensitivity, this effect can be removed by using an intensity-subtraction detection scheme.


Also related to this second feature there is a minor, but somewhat unusual, effect if the third mode is seeded with coherent light (the unseeded mode of the second non-linearity, mode C in Fig. \ref{fig:setup}). One might think that this light would only harm the sensitivity since it does not pass through the sample and is not correlated with light from the first crystal. However, there is a minor initial \emph{improvement} in the sensitivity, which is quickly ruined if the intensity of the coherent seed is increased much further. This effect is very small compared to the effect of seeding the other modes (a factor of improvement of about 2-3, compared to orders of magnitude for the other inputs), but it is there. Admittedly the reason for this is mysterious to us. We plan to investigate further in ongoing work.

Next we examine intensity-subtraction detection, written as $\hat{S}_{ij}=\hat{I}_{i}-\hat{I}_{j}$. This is a commonly-used technique in interferometry as intensity noise common to both beams is canceled, and better fringe visibility is obtained. A downside of previous SU(1,1) interferometry schemes is that ``direct detection'' (total intensity measurement) is needed \--- or failing that homodyne detection. In our hybrid device we have the advantageous detection setup of an SU(2) interferometer, with the scaling likely competitive with SU(1,1).   

Though the equations generated are very large, it is useful to present a simple case, the minimum detectable phase-shift squared for coherent light injection into mode $B$ and intensity difference subtraction between modes $B$ and $C$ with the probe phase (and all other phases) set to zero:  

\begin{align}
\Delta\phi_{\mathrm{min}}^2=&\frac{1}{4(1+\beta^{2})^{2}}\left[(1+\beta^{2})\coth^{2}r_{1}\right. \nonumber \\
& \left. +\coth^{2}r_{2}(\beta^{2}\csch^{2}r_{1}+\sech^{2}r_{1})\right. \nonumber \\
& \left. +(1+\beta^{2})(\tanh^{2}r_{1}-2)\right].
\end{align}

Similar scaling (converging in most reasonable limits) is obtained for the phase sensitivity when coherent light is injected into mode $A$. This means that if the sample is robust and the detectors are sensitive, sub-shot-noise-limit sensitivity at high brightness is possible without shining bright light onto the detectors. Note that for subtracted measurements the ``U-shape'' of the sensitivity as a function of $r_{2}$ vanishes, as $\coth^{2}(r_{2})\rightarrow 1$ as the gain gets large. This is likely because the second crystal produces un-correlated noise which is canceled out by the subtraction. Also, in the high gain limit the function further simplifies to $e^{-2r}/4(1+\beta^{2})$. Note that these scalings are most-often \emph{not} the optimal that can be obtained, as more judicious choices of phase can perform better. In our Comparisons section we make further note of this.

In order to get a more-intuitive understanding of the action of the device on the optical fields we can examine in more detail modes B and C before and after the final beam splitter in the case when mode B is seeded with a coherent state. 

Consider a single-mode state that has been both quadrature-squeezed and displaced (see Fig. \ref{quad}). Depending on the the relative phases of the displacement and the quadrature-squeezing operations, this state may also be \emph{photon-number} (amplitude) squeezed \--- with maximum number-squeezing occurring when the state is quadrature-squeezed along the same direction it is displaced in quadrature space (this would be the opposite case of what is displayed in purple in the figure). Such states have more-sharply-defined photon numbers (field amplitudes), but less-certain phases.

However we have the case of a \emph{two}-mode-squeezed state, therefore the squeezing occurs not in single-mode quadrature space but in superposition-mode quadrature space \--- where the quadrature operators are defined as the sum of the individual mode operators. 

In our case after the first non-linearity the field is displaced only in one mode but squeezed in superposition between both. After the second non-linearity there is also mutual squeezing between modes A and C, so at that point squeezing is shared both between C and A, and B and A, but \emph{not} between C and B.  

Since both non-linearities are pumped with phase-locked coherent beams, the phase difference in between them ($\phi$) determines to what degree the two squeezings shared by mode A are out of phase in quadrature space. Though modes B and C are not mutually squeezed before the beam splitter they are connected via mode A, which carries information about the value of $\phi$ in the form of how it is squeezed (including both magnitude and direction of squeezing) between both: A and B, and A and C.

Then, after the final beam splitter, due to the fact that is no longer possible, even in principle, to determine which non-linearity any photon came from, this mutual squeezing is transferred to a squeezing between modes B and C (with some residual squeezing left over in the other combinations, as controlled by the relative phases of all the beams). Now that the squeezing is between \emph{these} two modes \--- one of which is also displaced \--- the state after the beam splitter can be number-squeezed to an extent controlled by the phase $\phi$. Therefore photon-number/intensity/amplitude-difference detection between these modes can reveal the phase in a super-sensitive manner as we have shown quantitatively.  

\section{Comparisons}

In this section we wish to compare our setup with both the ``traditional'' boosted SU(1,1) non-linear interferometer, and the induced coherence setup in the spontaneous-parametric-down-conversion approximation.

\begin{figure}[h!]
\includegraphics[scale=1.1]{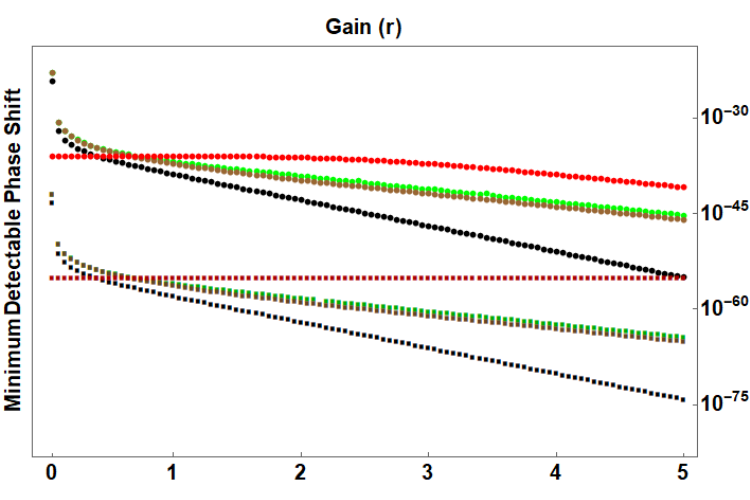}
\caption{The minimum detectable phase-shift squared of several ``fair comparison'' interferometric setups and detection schemes as a function of gain (of the first crystal for Mandel-type and of \emph{both} crystals for Yurke-type). Here we display the boosted Mandel-type setup with intensity detection at mode $B$ (green), intensity difference detection between modes $B$ and $C$ (brown), the boosted Yurke-type setup (black), and a standard coherent-light-seeded MZI with the extra light needed to create the aforementioned squeezings added to the initial input (red). The later is equivalent to the shot-noise limit. All other parameters are numerically optimized at each point. Note that this is not quite a true fair comparison as for the Yurke-type the gains of both non-linearities are set to the same value whereas in our scheme the gain of the second crystal is set to its (usually much lower) optimal value, giving the SU(1,1) an artificial advantage in this plot. The circular points (upper set) represent injected coherent light of about the same intensity as would be needed for a high-gain non-linearity, and the square points (lower set) represent a much-brighter coherent input. Note that the Heisenberg limit is not plotted as at ``fair comparison intensities'' it would be far below any others and not appear on the plot. At lab-available intensities for systems in which it could be realized, it would be far above. Coherent-light boosted schemes represent an intermediate regime where direct comparison to the HL is not a useful metric. \label{comapre2}} 
\end{figure}

To begin with the first point, it is instructive to examine the log plots of many different ``fair comparison'' setups. We take: our boosted Mandel-type setup (both intensity detection in one mode and intensity difference), the boosted Yurke-type setup \--- and a standard coherent-light-seeded MZI with the extra light that \emph{would be} needed to create the squeezing added to it. We compare the phase sensitivity of these three devices in a log plot in Figure \ref{comapre2}.

Though our setup does not win against the boosted Yurke-type, this is not quite a true fair comparison as for the Yurke-type the gains of both non-linearities are set to the same value, whereas in our scheme the gain of the second crystal is set to its (usually much lower) optimal value. We plan to investigate a ``more fair'' comparison in later work. The most important feature is that for all but very-low gains the boosted Mandel-type reaches orders of magnitude below the shot-noise (standard-quantum) limit. This is in addition to the other advantages present in the proposed device. 

It should be noted that at very low gains the phase sensitivity quickly decays. This is because at low gains the modes are only weakly-mixed \--- and since interferometry is based on mode-mixing, the power of the device in these regimes is limited in this case. 

Next we examine a similar setup but instead of general squeezing we take the ``first quantized'' approach and propagate an initial spontaneous-parametric-down-conversion (SPDC) state vector through the device up until before the final beam splitter (pre-beam-splitter), where it is

\begin{align}
|\psi_{p}\rangle=\frac{1}{\sqrt{2+\left|\beta\right|^{2}}}\hat{a}^{\dagger}_{p}\left(e^{i\phi}\hat{b}^{\dagger}_{p}+\hat{c}^{\dagger}_{p}\right)\left| 0,\beta ,0\right\rangle .
\end{align}

Since displacement and beam-splitter operators work in different representations/pictures it is necessary to then propagate detection operators back through the final beam splitter to work on the state above, with the standard transformations $\hat{b}^{\dagger}_{f}\rightarrow \frac{1}{\sqrt{2}}(\hat{b}^{\dagger}_{p}-i\hat{c}^{\dagger}_{p})$, etc. 

The full expression for the minimum detectable phase shift is large and so we do not report it here, but in the limit of large $|\beta|$ we find: $\Delta \phi_{\mathrm{min}}^2\rightarrow 19/4$. Thus, when taking the SPDC approximation the advantages of the induced-coherence interferometer are totally obscured. Since most mathematical treatments of the device make this assumption this is likely why the metrological power of this setup has gone mostly unnoticed.

\section{\label{sec:level3}Conclusion}

To conclude, we have theoretically investigated the induced-coherence interferometer when coherent seeds are added to all three input ports. The arrangement we study in the most detail is when the seed is injected into the initial arm which does \emph{not} pass through the phase-inducing sample.   

This scheme presents several practical advantages, which \--- strikingly \--- are available simultaneously with a single setup: super-sensitivity nearly on par with a boosted SU(1,1) interferometer, use of phase-stable intensity-subtraction measurements instead of additive (direct detection) or homodyning, the ability to use different frequencies for phase acquisition and detection (up into the infrared/terahertz regime), the favoring of an imbalance in the gains with \emph{low} gain in the second crystal, and the power to boost the sensitivity with a bright coherent seed which does \emph{not interact} with the sample to be studied. This last feature should in fact also be available to the boosted SU(1,1) setups \--- however to the best of our knowledge this fact has never been pointed out or pursued theoretically or experimentally. Furthermore, the same sensitivity scaling is observed when the mode with the phase is seeded with coherent light, even though this light does not reach the detectors. This last effect is not possible with other non-linear interferometers, and as far as we know is unique to the system we study here.

We conjecture that this scheme will be highly-advantageous in many real-world physical systems, especially those where high-quality phase estimation is desired for samples/detectors which are sensitive to bright lights (vulnerable to bleaching) and/or are interesting at wavelengths where good optical elements do not exist or are prohibitively expensive.\\

\section{Acknowledgments}

This paper is dedicated to the memory of Joe Haus, a gregarious and kind man, and an excellent scientist. He was an example to us all. It is also dedicated to the memory of Jon Dowling adviser to some of the authors. He always looked after his students with kindness and care, and made doing science exciting.

N. Miller would like to thank the U. Dayton Dean's Fellowship for funding, W. Plick thanks the U. Dayton SEED Grant, and S. Ramelow thanks the Deutsche Forschungsgemeinschaft (DFG) within the Emmy-Noether-Programm (RA 2842/1-1). Discussions with Z.Y. Ou were very useful.

\end{document}